# An Evolution Strategy Approach toward Rule-Set Generation for Network Intrusion Detection Systems (IDS)

**Herve Kabamba Mbikayi**

*Abstract*— **With the increasing number of intrusions in systems' and networks' infrastructures, Intrusion Detection Systems (IDS) have become an active area of research to develop reliable and effective solutions to detect and counter them. The use of Evolutionary Algorithms in IDS has proved its maturity over the times. Although most of the research works have been based on the use of genetic algorithms in IDS, this paper presents an approach toward the generation of rules for the identification of anomalous connections using evolution Strategies . The emphasis is given on how the problem can be modeled into ES primitives and how the fitness of the population can be evaluated in order to find the local optima, therefore resulting in optimal rules that can be used for detecting intrusions in intrusion detection systems..**

## I. INTRODUCTION

Intrusion Detection Systems (IDS) have become since many years an open area for research due the increasing amount of sophisticated intrusions and techniques developed by intruders to compromise security. An intrusion detection system is a device or software application that monitors network and/or system activities for malicious activities or policy violations and produces reports [10].

They are used as a countermeasure to preserve accessible infrastructures and to ensure data integrity against intrusions attempts [3]. A secure system or infrastructure might be considered as an infrastructure implementing correct design for its security to limit the probability of successful attacks. This latter one should possess some mechanisms to first detect probable attacks and take the related measures to prevent any damage. The detection of intrusions in a system or network is made possible by the use of IDS which are able to monitor the network or system activity to detect based on certain rules, suspected connections to report or neutralize.

An intrusion can be seen as actions or traffic not legally allowed on a system or network. These actions are performed by intruders. Generally an intruder is defined as a system, program or person who tries to and may become successful to break into an information system or perform an action not legally allowed [4]. Intruders are considered belonging to two categories, internal and external. The first one refers to a group of persons having legal access to a system and attempting to perform an unauthorized action. The second one refers to those who have no access to the system and try to use any means to perform actions within it or to compromise its security.

Intrusions refer to any set of actions that attempts to compromise integrity, confidentiality, or availability of a computer resource [4]. IDS systems are based on rules defined to match a particular traffic that can be considered as an intrusion. They use two categories of techniques to detect them: misuse detection and anomaly detection, which respectively refers to techniques that are characterized by known methods to penetrate a system (such as signatures, sequence of packets, etc.), and techniques that are defined and characterized by the identification of behaviors that deviate from the defined one [5]. Two categories of IDS exist: Host-based IDS and Network-based IDS. The first one refers to a system that monitors files, processes and activity related to a software environment associated to a particular host while the second one refers to a system deployed to monitor, and analyze network traffic to find any matching information based on the defined rules.

Based on expert knowledge, historical information on traffic or network activities, connections can be classified in two categories which are normal connection or intrusions. The IDS should be able from this information to generate rule-sets that separate normal connections from abnormal ones, thus producing information on how to detect intrusions on the system. This can be intuitively explained by the fact that a single rule cannot accurately classify all the attack types [2]. This process being a multi-variables optimization problem is suitable for evolutionary algorithms. However, most of the research works have applied Genetic Algorithms (GA) in the process. The use of GA implies finding correct representation of the chromosomes to suit the problem since they use binary representation. One question is raised however in the use of GA which put emphasis on the recombination operators and use mutation to extend the search space. The connections information or primitives such as IP addresses have some constraints such as the maximum value which reaches 255.255.255.255 [5]. The mutation operation is likely to produce offspring that overpass this constraint at high frequency rates.

Using ES which consider the impact of the genes in the phenotypes of individuals, connections' information can be represented from their phenotypes values and a control on them to meet the constraints can be applied easily in the evolution process.

This paper shows how ES can be used as a learning tool for IDS toward the generation of rule-sets for intrusion detection. The rest of the paper is organized as follows: In section 2, related works on the subject are discussed, section 3 gives an overview of ES, in section 4, we discuss about their application with IDS while in section 5 we draw some conclusion and talked about our future work.

## II. RELATED WORKS

In this section, we describe some important works related to the application of evolutionary algorithms in intrusion detection.

During the past days, many works addressing Intrusion Detection Systems issues proposed some approach based on EA to be used in many scenarios. Some of them was used to derive classification rules, other ones was used to detect intrusion in real-time while others one was used in parallel with others artificial intelligence techniques to derive mutated forms of attacks based on known information that can be launched on a system [11].

In [5], an approach of applying GA to detect intrusions was proposed. It included some quantitative features of network data for deriving classification. Categorical features have been also described in the work but no experimental results are available although the use of quantitative features may improve the detection rate.

In [11], a multi-objective approach with Evolution Programming (EP) to develop new method for detecting intrusions has been used. The work addressed particularly the problem of mutation or modification of attacks that can take another form based on previous known information. By vaccinating the system injecting some knowledge about an attack, using EP finite states (antibodies) transducers representing the population evolves to detect attacks.

In [2], a GA was applied to classify all types of smurfs attacks using a training dataset. Low false positive rate and high detection rate have been obtained.

In [12], based on historical data network on a dataset, GP was used to derive a set of classification in parallel with support-confidence framework as the fitness function. Several network intrusions were accurately classified with that approach.

## III. EVOLUTION STRATEGIES

Evolution Strategies are a sub-class of nature-inspired direct search methods belonging to the class of Evolutionary Algorithms (EA) which use mutation, recombination, and selection applied to a population of individuals containing candidate solutions in order to evolve iteratively better and better solutions. They were first introduced by Rechenberg [7], and Schwefel [8], for optimizing real-vectors [9]. ES are considered as other forms of EA and their concept is based on the biological evolution and adaptation. ES imitates, in contrast to GA, the effects of genes on the phenotype. They use a simple formalism to represent individual as real-valued vectors of object $x^{(t)} \in \mathcal{R}^n$.

The presumption for coding the variables in the ES is the realization of a sufficient strong causality (small changes of the cause must create small changes of the effect). The evolutionary progress takes place only within a very narrow band of the mutation step size which needs a clear definition of the rules that are used for the mutation step-size. During the optimization process, parameters of an ES are self-adapting and the emphasis is put on the mutation contrarily to GA which emphasize on the crossover operator.

In their use, they can be noted as (1+1)-ES to express the simplest form of ES in which one parent $x^{(t)}$ produces one offspring $x'^{(t+1)}$ by mutation, the offspring is assigned a fitness value evaluating its quality with respect to the problem, the offspring competes with the parent based on the fitness, and the better individual becomes the parent of the next generation. Thus, the selection of individuals for the next generation t+1 is deterministic [9].

$$x^{(t)} = \begin{cases} x'^{(t+1)} & if\ fi\left(x'^{(t+1)}\right) \geq fi\left(x^{(t)}\right) \\ x^{(t)} & otherwise \end{cases} \qquad (1)$$

Others formalisms exist to describe the ES such as (μ/ρ, λ)-ES, (1+λ)-ES, (1, λ)-ES. In the first notation, the symbol μ is used to represent to total number of parents while ρ represents the number of parents that will be recombined and, λ which stands for the number of offspring. The second notation describes an ES that used the "plus selection" in which the selection will happen among the parents and offspring. The latest one describes an ES using the "comma selection" in which the selection happens only among the offspring.

$$x'^{(t+1)} = x^{(t)} + \sigma^{(t)} \cdot N\ (0,1) \quad (2)$$

To perform the mutation operation on $x^{(t)}$, a random value $N\left(0, \sigma^{(t)}\right)$ from a normal distribution with a mean of zero and, using the 1/5th-rule which is based on the rate of successful mutations, the global step-size ($\sigma^{(t)} \in \mathcal{R}^+$ used during the mutation is itself adapted (Eq(2)). We consider successful mutation only when the produced offspring dominates its parent with respect to the fitness expressed as $x'^{(t+1)} \geq x^{(t)}$ for a function to maximize.

$$\sigma^{(t+1)} = \begin{cases} \sigma^{(t)} \cdot \alpha_{ES} & if\ fi(x'^{(t+1)}) \geq fi(x^{(t)}) \\ \sigma_{ES}^{-\frac{1}{4}} & otherwise \end{cases} \qquad (3)$$

$N\ (0,1)$ represents the normal distribution with mean of zero and standard deviation of one. $\alpha_{ES}$ is the change rate of the global step-size. Values for $\alpha_{ES}$ are recommended between $2^{1/n}$ and 2, where $n$ is the dimension of the problem. Each element $x_i^{(t)}$ is initialized to a value $x_i^{(0)}$ and $\alpha^{(t)}$ is initialized to a constant value and $\alpha^{(0)}$ the value for this constant depends on the problem. $\alpha_{ES}$ represents the changing rate of the step-size [9].

## IV. EVOLUTION STRATEGY AND INTRUSION DETECTION SYSTEMS

The use of EC and artificial intelligence techniques in IDS has played a major role in the intrusion detection. Most of the works focused on the application of GA in intrusion detection to generate optimal rule-sets and to detect intrusions. ES are powerful optimization tools that showed their maturity in many fields during decades and nowadays. Contrarily to GA which are based on the abstraction of the problem to chromosome binary representation, ES emphasize on the effects of the genes to the phenotypes of individual. They use real-valued representation instead of binary as do GA.

With their particularity in working with numbers, ES can be used to generate rule-sets for IDS that separate normal connections on the network from anomalous ones. IDS need to have some stored rules that enable them to detect intrusions. These rules are generally presented in the following form:

*if {condition} then { act }*

The condition refers the event in which a particular connection matches a rule in the IDS. The "act" can be seen as an action that needs to be performed based on the detected intrusion. A condition can be i.e. a blacklisted source IP connecting to a mail server and the action can be to generate an alert, sending messages, or stopping the connection, etc. An example of a rule may be:

*if { the source IP 213.255.212.10; destination 78.0.0.1, on the port 25; and the connection lasts at least 8 seconds} then {stop the connection}*

The above example is a scenario of a rule that will stop each connection originating from the IP address 213.255.212.10 to the mail server (destination IP 78.0.0.1 and port 25) if this connection lasts at least 8 seconds. This source IP address may have been added to a list of blacklisted addresses.

In this paper, we use ES to generate such rules that will match only anomalous connections. These rules are tested on historical data and then used to detect new suspicious connections. Our approach is based on the use of pre-classified connections at an expert level producing a dataset that differentiates normal connections from anomalous ones. This information is obtained using a sniffer (such as SNORT, TcpDump, etc.). To evaluate our population during the execution of the ES algorithm, this dataset is used to determine the adaption level of population

Table1. Rule definition for connection and range of values of each field

| Attribute | Range of values | Description | Example |
|---|---|---|---|
| Source IP address | 0.0.0.0~255.255.255.255 | 192.168.17.0/24 | A network in the range 192.168.17.0~192.168.0.255 |
| Destination IP address | 0.0.0.0~255.255.255.255 | 10.1.0.0/16 | A network in the range 10.1.0.0~10.1.255.255 |
| Source port | 0~65535 | 1245 | Source of port the connection |
| Destination port | 0~65535 | 25 | Destination port indicating a mail server connection |
| Duration | 0~99999999 | 612 | Duration of the connection is 612 seconds |
| State | 1~20 | 11 | The connection is terminated by the source for internal use |
| Protocol | 1~9 | 2 | The protocol used is TCP |
| Number of bytes sent by sources | 0~9999999999 | 8645 | The source sent 8645 bytes of data |
| Number of bytes sent by responder | 0~9999999999 | 68458 | The responder sent 68458 bytes of data |

*A. Problem abstraction to ES primitives*

Primitives in ES are represented by vectors of real numbers. Table 1 describes the information needed by IDS to create rule-set that will be used to match anomalous traffic. From this latter one, the representation of all information is based on numbers. Since the ES act on phenotype expression of the individuals, it is possible to derive from this information, an abstraction of the problem we want to optimize without any need to model it as a particular string or binary representation. We just need to express all the information in their respective decimal representation; we then find the constraints of the ES that apply to this particular domain. For our ES, we chose to compose the rule on the IDS using only six attributes which are the source range of IP addresses, the destination range of IP addresses and the destination ports range expressing the scenario of most common rule's definition in network security appliances.

Table 2: Example of a connection's attribute in decimal values

| Source IP | Destination IP | Destination Port |
|---|---|---|
| 1360267777 | 1678445057 | 53 |

Table 2 shows an example of decimal values of some connection attributes (IP source 81.20.10.1, IP destination 100.11.10.1 and port 53) that are used as primitives of our ES. The ES will deal directly with the phenotypes of the connection attributes expressed in decimal values (integer) and during the evolution process they will be converted into real numbers by the application of ES operators.

### B. Generation of the population

The first step for an evolutionary algorithm is to generate random population which is a set of rules that will evolve throughout the evolution process toward optimal values.

The ES will use the pre-classified dataset of connections to evaluate the fitness of the population that competes for survival. A rule contained in the population can be expressed as:

*If { Source IP in the range 1360267777~186026777, Destination IP 1678445057~ 1678445057 and destination port 1200~3150} then { Stop the connection}*

To generate the population, we take into account the constraints of the search space of our domain which are that IPv4 addresses exist only on the range of 0.0.0.0 to 255.255.255.255. This information can be converted in decimal values representation as 0 to 4294967295 and constraints on the ports as shown in Table 1 should also be taken into consideration. Ports are generated on the range of 0~65535.

### C. Evaluation function

During the evolution process, generated rules are recombined, mutated and evaluated to find the accurate rule set for the IDS.

$$H = \frac{y_1}{x_1} + \frac{x_2}{y_1} + \frac{y_2}{x_3} + \frac{x_4}{y_2} + \frac{y_3}{x_5} + \frac{x_6}{y_3} \quad (4)$$

In Eq. 4 we calculate the distance between each primitive by dividing the greatest number by the smallest one and summing them to find the total distance. This is very important for the optimization of the number of rules that will be used in the IDS. The greater the distance, the greater the probability that many connections will match a single rule to avoid individual rules for each connection.

In the dataset that contains pre-classified connections, each rule primitives $(x_i)$ is compared individually to the connections primitives $(y_i)$ present in the dataset. Since we use six primitives with ES to generate rule-set, for a rule to match a connection in the dataset, the following constraints must be met (Eq. 5):

$$\begin{aligned} x_1 &\leq y_1 \leq x_2 \\ x_3 &\leq y_2 \leq x_4 \\ x_5 &\leq y_3 \leq x_6 \end{aligned} \quad (5)$$

Each rule is evaluated individually to see if it matches a connection on the pre-classified dataset. If it matches a normal connection then the value "0" is added to the "match" number and if it matches a suspicious connection, the value "1" is however added.

$$fitness = match * H \quad (6)$$

The number of matching is then multiplied by the total distance to find the fitness (Eq. 6). Note that, we are maximizing the fitness value during the optimization.

### D. Evolution Strategy operators

Recombination is applied during the evolution process to produce one child by averaging the value of the parents. However we should be careful in the implementation to avoid having many values that exceed the range of connection attributes (IP address, Ports). The mutation is applied by adding random noise drawn from normal distribution as expressed in Eq. 2, and then the selection is made upon the best individuals in the population.

The rules are then selected and updated in Rule Base of the IDS. The performance of ES largely depends on the adjustment of its parameters, prominently the mutation strength(s) [6]. Important decision must be taken at this level on it value by observation and by referring to the given problem.

### E. System Architecture

Data are collected using a sniffer program (Tcpdump), loading a binary dataset file from DARPA, they are then analyzed at an Expert level to differentiate good connections from anomalous ones. The expert is part of the system of the system since its contributions is a must to build the evaluation function that the ES will use. The identified normal and abnormal connections results are then used by the ES to evaluate the generated rules and produce better one to update the Rule Base.

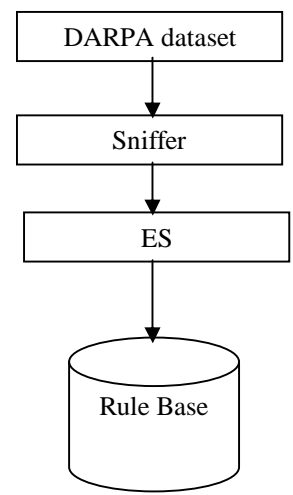

Figure 1: System architecture

## V. CONCLUSION

This paper presented a way to generate rule-sets using ES, which stand for powerful tools for optimization and which can be considered to be suitable for optimizing rule-set in IDS. An overview of their application with IDS have been presented. Emphasizing on the way the problem can be abstracted to suit ES primitives, we identified the constraints for the search space of the connection attributes such IP addresses, Port numbers.

In our future work we would like to investigate further on the way additional attribute can be included in the chromosome, such as temporal information while it is believed that a profound study on the ES application to IDS would lead us to the use of multi-objective ES [1], to principally reduce the number of rules generated by optimizing both the matching chromosomes and at the other side optimizing the number of matching connections in the dataset.

as a Consultant in security, Manager and Director of Information and Technology departments.

His more recent works was as Head of IT department at MONEYTRANS Group international, Technical Director at Advanced Center for Information and Technology (ACIT) based in Congo.

He also possess many IT vendors' certifications including Cisco Certified Network Associate (CCNA), Cisco certified Network Professional (CCNP), Cisco certified Pix and ASA, etc.

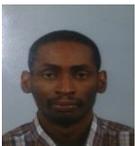

**Herve Kabamba Mbikayi**
Guest Reasearcher at Leiden Institute of Advanced Computer Science (LIACS), Leiden University, The Netherlands
Research and Teaching Assistant at Institut Superieur de Commerce de Kinshasa, Kinshasa, Dem. Rep. of Congo

Author of a previous paper titled "**Visual Composition and Automatic Code Generation for Heterogeneous Components Coordination with Reo**" accepted for publication in the International Journal of Advanced Research in Computer Science and Electronics Engineering, Volume 1, Issue 1, March 2012, he is member of the Leiden embedded System Research Group (LERC) of the Leiden University . He got his  Bac+ 5 graduate diploma in computer science from the University of Kinshasa since  2007 and had been working in the IT industry for more than 10 years